\documentclass[pra,twocolumn,superscriptaddress,floatfix]{revtex4-1}
\usepackage{mathrsfs}
\usepackage{amsfonts}
\usepackage{amsmath}
\usepackage{mathtools}
\usepackage{txfonts}
\usepackage{xcolor}
\usepackage{units}
\usepackage{amssymb}
\usepackage{graphicx}
\usepackage{bm}
\usepackage{color}
\usepackage[normalem]{ulem}
\usepackage[caption=false]{subfig}

\newcommand{\ket}[1]{\left|#1\right\rangle}
\newcommand{\bra}[1]{\left\langle #1\right|}

\begin{document}

\title{Universal quantum computation and quantum error correction using discrete holonomies}
\author{Cornelis J. G. Mommers}
\affiliation{Department of Physics and Astronomy, Uppsala University,
Box 516, Se-751 20 Uppsala, Sweden}
\author{Erik Sj\"oqvist}
\email{erik.sjoqvist@physics.uu.se}
\affiliation{Department of Physics and Astronomy, Uppsala University,
Box 516, Se-751 20 Uppsala, Sweden}
\date{\today}
\begin{abstract}
Holonomic quantum computation exploits a quantum state's non-trivial, matrix-valued 
geometric phase (holonomy) to perform fault-tolerant computation. Holonomies arising 
from systems where the Hamiltonian traces a continuous path through parameter space 
have been well-researched. Discrete holonomies, on the other hand, where the state 
jumps from point to point in state space, have had little prior investigation. Using a 
sequence of incomplete projective measurements of the spin operator, we build an 
explicit approach to universal quantum computation. We show that quantum error correction 
codes integrate naturally in our scheme, providing a model for measurement-based quantum 
computation that combines the passive error resilience of holonomic quantum computation 
and active error correction techniques. In the limit of dense measurements 
we recover known continuous-path holonomies.
\end{abstract}

\maketitle

\section{Introduction}
In the circuit model of quantum computation, information is processed by using a series 
of quantum gates on a register of qubits. These gates are unitary transformations, which 
can be realised using non-Abelian geometric phases (holonomies) \cite{zanardi99}
that make them intrinsically fault-tolerant \cite{pachos01}. Holonomic quantum computation is 
often studied in adiabatic systems, where the Hamiltonian traces a continuous path 
through parameter space. Experimental realisations of adiabatic holonomic quantum 
computation in a wide range of physical settings have been proposed 
\cite{duan01, faoro03, solinas03} and implemented \cite{toyoda13,leroux18}. On the other 
hand, there has not been as much research into the discrete case, whereby the state jumps 
from point to point in state space resulting in non-Abelian holonomies 
\cite{anandan89,sjoqvist06,oi14}. 

Here, we examine holonomies arising from sequences of incomplete measurements. 
We demonstrate a universal set of holonomic quantum gates acting on qubits implemented 
by a discrete set of incomplete measurements on spin coherent states (SCSs) 
\cite{radcliffe71,peres95}. SCSs can be created in the laboratory \cite{estrada2013,chalopin2018} 
and have mathematically desirable properties \cite{chryssomalakos18}. In the limit of 
dense measurements 
our scheme reduces to well-known continuous-path holonomies, which, for SCSs, are Abelian 
\cite{zee88} and therefore insufficient to achieve universality. Hence, to realise truly 
non-Abelian holonomies acting on SCS-qubits it is essential to use (discrete) sequences 
consisting of a finite number of measurements. 

We further show that measurement-driven holonomies on SCSs can be naturally 
merged with active error-correcting techniques, such as direct implementation of the bit 
flip repetition code \cite{peres85} and extension into the nine-qubit Shor code \cite{shor95}. 
In this way, our proposed scheme can be viewed as a model for measurement-based 
quantum computation, in the same vein as, e.g., one-way cluster state quantum computation 
\cite{raussendorf01} and teleportation-based quantum computation \cite{nielsen03}. 
Our approach combines the passive error resilience of holonomic quantum computation and 
active error correction techniques making it a promising tool for robust quantum computation.  

\section{Holonomic scheme}
\subsection{Preliminaries}
To introduce notation, we begin by briefly outlining some conceptual aspects of discrete 
holonomies. Suppose we have an $N$-dimensional Hilbert space. A projection onto a 
$K$-dimensional subspace $p_a$, spanned by a (nonunique) frame 
$\mathcal{F}_a = \{ \ket{a_k} \}_{k=1}^K$, can be realised with a projector $P_a$. The 
overlap matrix, defined component-wise as 
    \begin{eqnarray}
        \left( \mathcal{F}_a | \mathcal{F}_b \right)_{kl} = \langle a_k \ket{b_l},
    \end{eqnarray}
quantifies how different subspaces are connected \cite{mead91}. From a cyclic sequence 
$\mathcal{C}$ of $q+1$ projections ($q$ of which are distinct) we 
can construct $\Gamma_{\mathcal{C}}$, given by
    \begin{eqnarray}
        \Gamma_\mathcal{C} = P_1 P_q P_{q-1}\cdots P_1.
    \end{eqnarray}
$\Gamma_\mathcal{C}$ can be realised by applying a sequence of $q+1$ projective 
filtering measurements that transforms an input state $\ket{\psi} \in p_1$ as 
\begin{eqnarray}
\ket{\psi} \to \sum_{k,l} \ket{1_k} D_{kl} \bra{1_l} \psi \rangle, 
\end{eqnarray}
with $D = \left( \mathcal{F}_1 | \mathcal{F}_q \right) \left( \mathcal{F}_q | \mathcal{F}_{q-1} 
\right) \cdots \left( \mathcal{F}_2 | \mathcal{F}_1 \right)$ \cite{sjoqvist06}. The holonomy of 
$\mathcal{C}$ is the unitary part $U_D = \left| D \right|^{-1} D$ of $D$, with 
$|D| = \sqrt{DD^\dagger}$ the positive part of $D$. In the following, 
we take $p_1$ to be the computational subspace.  

Next, let us parametrise a unit vector $\bf{n}$ with spherical 
coordinates as $\left( \sin \theta \cos \phi, \sin \theta \sin \phi, \cos \theta \right)$. With 
$\left\{ \ket{j,m} \right\}_{m=-j}^j$ denoting the eigenkets of the $J_z$ operator, the SCSs read 
$e^{-i \phi J_z} e^{-i \theta J_y} \ket{j,\pm j}$ (we put $\hbar = 1$ from now on, and abbreviate 
$\ket{j,\pm j}$ to $\ket{\pm j}$). If $j \geq 1$,  
we can have our sequence of frames be projections onto 
subspaces spanned by different SCSs, viz., 
    \begin{eqnarray}
        \mathcal{F}_a\left( \theta_a,\phi_a \right) = \left\{ e^{-i \phi_a J_z} 
        e^{-i \theta_a J_y} \ket{j,\pm j} \right\} = \left\{\ket{\pm j ;  {\bf{n}}_a} \right\} .
    \end{eqnarray}
Each projective filtering measurement corresponds to the operator 
$P_a = \ket{+ j ; {\bf{n}}_a} \bra{+ j ; {\bf{n}}_a} + \ket{- j ; {\bf{n}}_a} \bra{- j ; {\bf{n}}_a}$ 
and represents the degenerate measurement outcome $j^2$ of the observable 
$\left( {\bf n}_a \cdot {\bf J} \right)^2$.

We can calculate the overlap matrix for the SCSs by decomposing each state 
into a tensor product of spin-$\frac{1}{2}$ states, $\ket{\pm j} = \ket{\pm \frac{1}{2}}^{\otimes 2j}$. 
One finds nonvanishing overlap matrices \cite{sjoqvist06}
    \begin{eqnarray}
        \left( \mathcal{F}_a | \mathcal{F}_b \right) =  \left( \begin{array}{cc} 
        R_{a,b} & S_{a,b} \\ (-1)^{2j} S_{a,b}^{\ast} & R_{a,b}^{\ast} 
        \end{array} \right), 
    \end{eqnarray}
with
    \begin{eqnarray}
        R_{a,b} & = & \left[ \cos \left( \frac{\theta_a-\theta_b}{2} \right) \cos \left( \frac{\phi_a-\phi_b}{2} 
        \right) \right. 
        \nonumber \\ 
        & & \left. + i \cos \left( \frac{\theta_a+\theta_b}{2} \right) \sin \left( \frac{\phi_a-\phi_b}{2} \right) 
        \right]^{2j}, 
        \nonumber \\  
        S_{a,b} & = & \left[ \sin \left( \frac{\theta_a-\theta_b}{2} \right) \cos \left( \frac{\phi_a-\phi_b}{2} 
        \right) \right. 
        \nonumber \\ 
        & & \left. - i \sin \left( \frac{\theta_a+\theta_b}{2} \right)  \sin \left( \frac{\phi_a-\phi_b}{2} \right) 
        \right]^{2j}. 
    \end{eqnarray}

To make the holonomy unambiguously associated with a quantum gate we require that the 
overlap matrices are unitary up to a multiplicative positive number. From now on we 
restrict $j$ to $j = (2n+1)/2$, $n \in \mathbb{N}$. Then all the SCSs subspaces will be fully 
overlapping \cite{kult06}. We define our qubit as $\ket{0} = \ket{j}$ 
and $\ket{1} = \ket{-j}$. Our input state is $\ket{\psi} = a\ket{j} + b\ket{-j}$ with 
$\left|a\right|^2+\left| b \right|^2=1$ and $\ket{\pm j} \in \mathcal{F}_1$, where 
$\mathcal{F}_1$ spans the first subspace. A $\ket{j}$ state will be composed of 
$\left( 2n + 1 \right)$ spin-$\frac{1}{2}$ constituents \cite{remark0}. We can make 
a left polar decomposition of the overlap matrix into
    \begin{eqnarray} 
        \left( \mathcal{F}_a | \mathcal{F}_b \right) = \left| \left( \mathcal{F}_a | 
        \mathcal{F}_b \right) \right| U_{a,b} = \kappa_{a,b}^{-1} U_{a,b},
    \end{eqnarray}
where $\kappa_{a,b}^{-1} = \sqrt{\left| R_{a,b} \right|^2 + \left| S_{a,b} \right|^2} >0$ and 
$U_{a,b}$ is a unique unitary matrix \cite{remark1}. If we perform a sequence of $q+1$ 
measurements projecting onto the SCS subspaces, the final holonomy becomes
    \begin{eqnarray}
        U_D & = & \kappa_{1,q} \kappa_{q,q-1} \cdots \kappa_{2,1} \nonumber \\ 
         & & \times \left( \mathcal{F}_1 | \mathcal{F}_q \right) \left( \mathcal{F}_q | 
         \mathcal{F}_{q-1} \right) \cdots \left( \mathcal{F}_2 | \mathcal{F}_1 \right). \label{eq:hol}
    \end{eqnarray}

\subsection{Single-qubit gates}
We consider sequences of four measurements, which is the minimum number of measurements 
that can yield a non-trivial holonomy \cite{remark1.5}. The first measurement is a projection 
onto the same subspace as $\ket{\psi}$. This projection is the preparation of our state 
in the correct subspace. We then carry out two more projections onto subspaces that are 
different from the first subspace and each other. Finally, we project back onto the same 
subspace as the starting measurement. Without loss of generality we define 
$\left( \theta_1,\phi_1 \right) = \left( \theta_4,\phi_4 \right) = \left( 0,0 \right)$. 
Our sequence of measurements becomes
$(0,0) \to \left( \theta_2,\phi_2 \right) \to \left( \theta_3,\phi_3 \right) \to (0,0)$.

An astute reader may raise objections to our use of $(0,0)$ as our starting state because 
spherical coordinates are undefined at the poles. There, we have $e^{-i\phi J_z}\ket{\pm j}$ 
and $e^{-i\phi J_z} e^{-i \pi J_y} \ket{\pm j}$, respectively, so there is no unique eigenstate. 
However, we were careful to define the $(0,0)$ label to correspond to our starting and ending 
subspace, which is unambiguous \cite{remark2}. Subsequent measurements are projections 
onto rotated versions of the original subspace, which is readily seen when we recall that 
$e^{-i \phi {\bf n} \cdot {\bf J}} = \mathcal{D}_{{\bf n}} (\phi)$ is the rotation operator around 
the direction ${\bf n}$.

We want to find the rotation gates, as any 
single-qubit operation can be decomposed exactly into $\mathcal{D}_z (\alpha) 
\mathcal{D}_y (\beta) \mathcal{D}_z (\gamma)$. Note that for 
the rest of this analysis any identification should be understood as an identification up to 
an unimportant global phase factor.

We first implement a rotation gate about the $z$ axis. We pick $\left( \theta_2,\phi_2,\theta_3,
\phi_3 \right) = \left( \frac{\pi}{2},\pi,\frac{\pi}{2},\varphi \right)$, where $\varphi \in \left[ 0,2\pi \right)$ 
and find 
    \begin{eqnarray}\label{eq:rotz}
         \mathcal{D}_z (\phi) = U_{D,z} & = & \frac{1}{\sqrt{\cos^{2+4n}\left( \frac{\varphi}{2} \right) + 
         \sin^{2+ 4n} \left( \frac{\varphi}{2} \right)}} \left( \begin{array}{cc} z^{\ast} & 0 \\ 0 & z 
         \end{array} \right) 
         \nonumber \\ 
        & = & \left( \begin{array}{cc} e^{-i\phi/2} & 0 \\ 0 & e^{i\phi/2} \end{array} \right), 
        \label{eq:phase}
    \end{eqnarray}
with
    \begin{eqnarray}
        z = e^{i (2n+1) \frac{\varphi}{2}} \left( (-1)^n \cos^{2n+1} 
        \left(\frac{\varphi}{2} \right) - i \sin^{2n+1} \left(\frac{\varphi}{2} \right) \right),
    \end{eqnarray}
where the final identification of Eq.~(\ref{eq:rotz}) holds since $U_D \in \mathrm{SU}(2)$, 
so $|z| = 1$. The relative phase change induced by this holonomy is 
    \begin{eqnarray}\label{eq:anglephase}
        \phi = \arg (z/z^{\ast}) \in (-\pi,\pi] . 
    \end{eqnarray}  
Rotations about the $x$ and $y$ axes are given by 
    \begin{eqnarray}
        U_{D,x} & = &  \frac{1}{\sqrt{\cos^{2+4n} \left( \frac{\varphi}{2} \right) + 
        \sin^{2+ 4n} \left( \frac{\varphi}{2} \right)}} \left( \begin{array}{cc} {\rm Re}(z) & 
        i{\rm Im}(z) \\ i{\rm Im}(z) & {\rm Re}(z)) \end{array} \right) , \nonumber \\
        U_{D,y} & = & \frac{1}{\sqrt{\cos^{2+4n} \left( \frac{\varphi}{2} \right) + \sin^{2+ 4n} 
        \left( \frac{\varphi}{2} \right)}} \left( \begin{array}{cc} {\rm Re}(z) & -{\rm Im}(z) \\ {\rm Im}(z) & 
        {\rm Re}(z)) \end{array} \right) .  
    \end{eqnarray}
These matrices are equivalent to the following list of angles:
    \begin{eqnarray}
        U_{D,x} = \mathcal{D}_x (\phi) : (0,0)  & \to &  \left( \frac{\pi}{2},\pi \right) \to 
        \left( \varphi,\frac{\pi}{2} \right) \to (0,0) , 
        \nonumber \\
        U_{D,y} = \mathcal{D}_y (\phi) : (0,0)  & \to &  \left( \varphi, \begin{array}{c} 0 \ 
        {\rm if} \ n \ {\rm is \ even} \\ \pi \ {\rm if} \ n \ {\rm is \ odd} \end{array} \right) 
        \nonumber \\ 
         & \to & \left( \frac{\pi}{2},\frac{\pi}{2} \right) \to (0,0) ,
    \end{eqnarray}
where $\varphi$ can be found by solving Eq.~(\ref{eq:anglephase}). 

\subsection{Two-qubit gate}
To construct an entangling two-qubit gate necessary for universality we make use of the 
auxiliary states $\ket{\zeta_\pm} \in \mathcal{F}_1$ for the second qubit, with $\langle \zeta_+ 
\vert \zeta_- \rangle = \xi$, in addition to the input state $\ket{\psi}$ of the first qubit. Our 
two-qubit gate only rotates the first qubit, keeping the subspace of the second qubit fixed. 
In the extended Hilbert space our new frames are $\mathcal{F}_a = \left\{\ket{\pm j ; {\bf{n}}_a }
\otimes \ket{\xi_\pm} \right\}$, with corresponding rank-two projection operators 
$P_a = \ket{+j ; {\bf{n}}_a}\bra{+ j ; {\bf{n}}_a }\otimes \ket{\xi_+}\bra{\xi_+} + \ket{- j ; 
{\bf{n}}_a}\bra{- j ; {\bf{n}}_a} \otimes 
\ket{\xi_-}\bra{\xi_-}$. The new overlap matrices are 
    \begin{eqnarray}
        \left( \mathcal{F}_a | \mathcal{F}_b \right) =  \left( \begin{array}{cc} 
        R_{a,b} & \xi S_{a,b} \\ - \xi^\ast S_{a,b}^{\ast} & R_{a,b}^{\ast} 
        \end{array} \right) = \sqrt{\left| R_{a,b} \right|^2 + \left| \xi S_{a,b} \right|^2} \tilde{U}_{a,b}. 
    \end{eqnarray}
If $\ket{\zeta_+} = \ket{\zeta_-}$, then $\xi = 1$ and we recover the single-qubit gates 
acting on the first qubit. If $\xi = 0$, then 
    \begin{eqnarray}
        \left( \mathcal{F}_a | \mathcal{F}_b \right) =  \left( \begin{array}{cc} 
        R_{a,b} & 0 \\ 0 & R_{a,b}^{\ast} \end{array} \right) = \left| R_{a,b} \right| \left( \begin{array}{cc} 
       e^{i\phi_{a,b}} & 0 \\ 0 & e^{-i\phi_{a,b}} \end{array} \right), 
    \end{eqnarray}
which is a phase gate inducing a phase shift $\Phi = \phi_{1,q} + \phi_{q,q-1} + \dots + \phi_{2,1}$. 
If we have an input state $\ket{\psi} \otimes \ket{\tilde{\psi}}$, where $ \ket{\tilde{\psi}} = 
c \ket{j} + d \ket{-j}$, with $\left| c \right|^2+\left| d \right|^2=1$, then the action of the gate is 
    \begin{eqnarray}
        & &\left( a\ket{j} + b \ket{-j} \right) \otimes \ket{\tilde{\psi}}  
        \nonumber \\
        & & \mapsto a e^{i\Phi} \langle \zeta_+ \vert \tilde{\psi} \rangle \ket{j} \ket{\xi_+} + b e^{-i\Phi} 
        \langle \zeta_- \vert \tilde{\psi} \rangle \ket{-j} \ket{\xi_-}. 
    \end{eqnarray}
This gate can entangle. For instance, if we pick $\ket{\zeta_\pm} = \ket{\pm j}$ the 
concurrence \cite{wootters98} will be proportional to $\left|a b c d \right|$, which is nonzero 
in general. The fact that we can construct all single-qubit gates from the rotation gates and 
that we can create an entangling two-qubit gate implies universality \cite{bremner02}.

\subsection{Example}
\begin{figure}[htb]
        \centering
        \includegraphics[width= 7 cm]{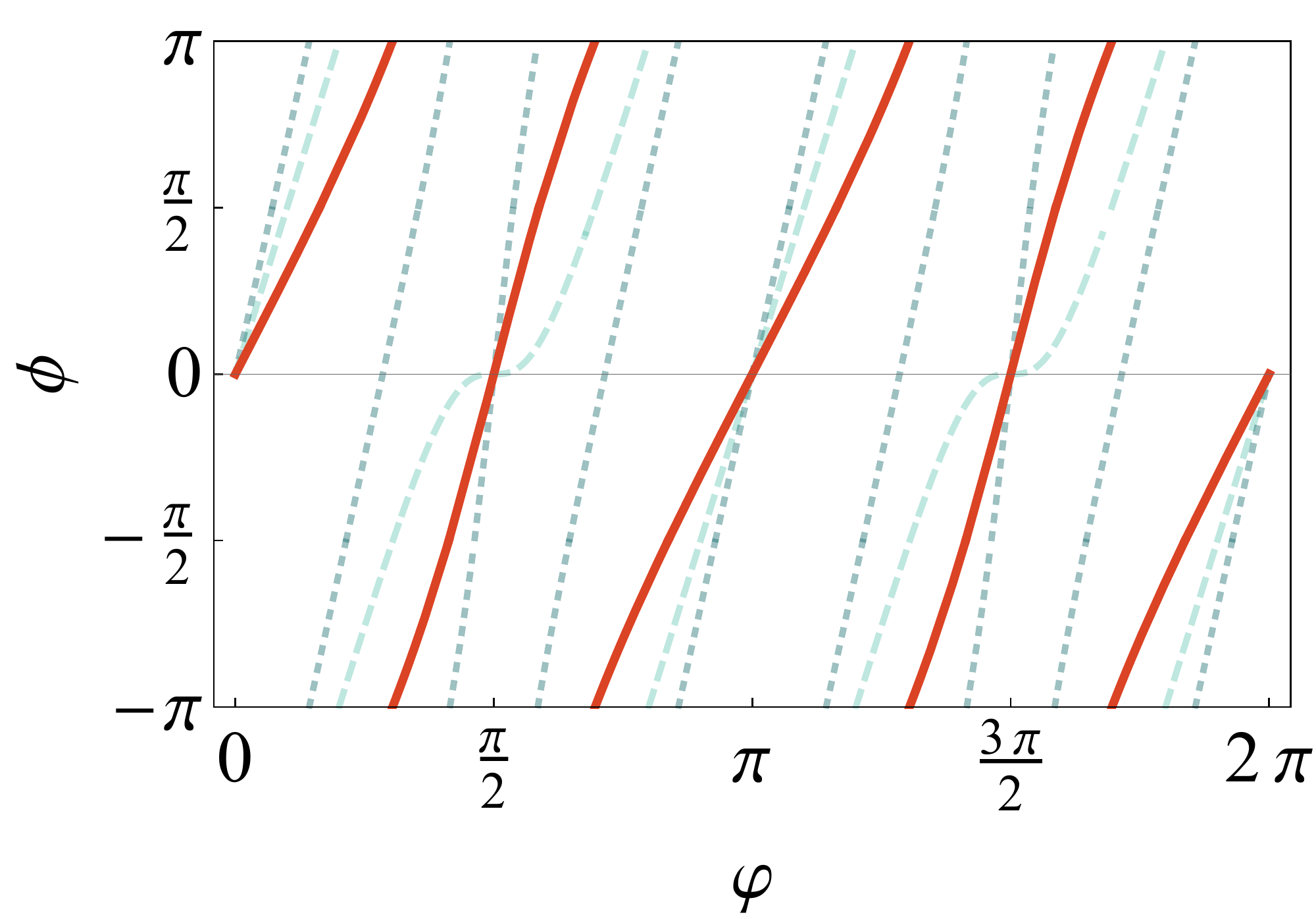}
        \caption{The relative phase $\phi$ as a function of the subspace rotation angle 
        $\varphi$ for $j=\frac{3}{2}$ (solid), $j = \frac{5}{2}$ (dashed) and $j = \frac{7}{2}$ 
        (dotted).}
        \label{fig:phase}
    \end{figure}

It is instructive to work out the discrete set of single-qubit Clifford + T gates \cite{giles13} 
(the S, H, and T gates) for $j = \frac{3}{2}$, which are sufficient for universal quantum computation 
when supplemented by an entangling two-qubit gate. The 
procedure for higher-$j$ gates is identical. Plugging in $n = 1$ into Eq.~(\ref{eq:phase}) gives us
    \begin{eqnarray}
        U_{D,z}^{(j=\frac{3}{2})} = \frac{1}{\sqrt{10 + 6 \cos ( 2\varphi )}} 
        \left( \begin{array}{cc} 1+3e^{-2i\varphi} & 0 \\ 0 & 1 + 3e^{2i\varphi} \end{array} \right).
    \end{eqnarray}
Therefore,
    \begin{eqnarray}\label{eq:phasej32}
        \phi = \arg \left( \frac{1+3e^{-2i\varphi}}{1+3e^{2i\varphi}} \right).
    \end{eqnarray}
A plot of Eq.~(\ref{eq:phasej32}) is shown in Fig. \ref{fig:phase}. Note that $\phi$ is multivalued. 
However, each intersection point yields the same holonomy. 
For the T and S gates we want to find the intersection point where $\phi$ is 
$\frac{\pi}{4}$ and $\frac{\pi}{2}$, respectively. H can be decomposed into a $\frac{\pi}{2}$ 
rotation around the $y$ axis and a $\pi$ rotation around the $z$ axis. We find that the gates 
can be realised using the following sequence of measurements:
\begin{eqnarray}\label{eq:angles}
{\rm T} : (0,0) & \to & \left( \frac{\pi}{2},\pi \right) \to \left( \frac{\pi}{2}, \phi_{\rm T} \right)  
\to (0,0), 
\nonumber \\
{\rm S}  : (0,0) & \to & \left( \frac{\pi}{2},\pi \right) \to \left( \frac{\pi}{2},  \phi_{\rm S} \right) 
\to (0,0),           
\nonumber \\
{\rm H} :  (0,0) & \to & \left( \phi_{\rm S},\pi \right) \to  
\left( \frac{\pi}{2},\frac{\pi}{2} \right) \to (0,0) \to \left( \frac{\pi}{2},\pi \right) 
\nonumber \\ 
 & \to & \left( \frac{\pi}{2},\arctan \sqrt{2} \approx 0.955 \right) \to (0,0),  
\end{eqnarray}
with $\phi_{\rm T} = 2 {\rm arcsec} \left(2 \sqrt{\frac{6}{\sqrt{6 \left(\sqrt{2}-\sqrt{36 
\sqrt{2}+70}+10\right)}+12}} \right) \approx 1.44$ and $\phi_{\rm S} = \arctan 
\left( \frac{3+\sqrt{17}}{2} \right) \approx 1.30$. Note that these sequences are not 
unique. For example, for the H gate we can replace the fifth pair of angles by 
$\left( \frac{\pi}{2},0 \right)$. 

\section{Gate performance and error-correction codes}  
Having constructed the formalism to describe holonomic quantum gates, we now turn to 
their performance. For each projection onto a subspace $p$ there is a chance our state 
collapses onto the orthogonal subspace $p^\perp$. So, we can quantify the gates' 
performance by looking at the total transition amplitude into the intended subspace. 
We focus on the rotation gates since they form the basis for all single-qubit gates. Furthermore, 
due to symmetry, we only need to look at rotations around the $z$ axis. It turns out that 
the transition amplitude $T_\mathcal{C}$ is independent of the input state and is given by
    \begin{eqnarray}
        T_\mathcal{C} = 4^{-2n} \left( \cos^{2+4n} \left( \frac{\varphi}{2} \right) +  \sin^{2+4n} 
        \left( \frac{\varphi}{2} \right) \right).
     \end{eqnarray}
For $j=\frac{3}{2}$ the maximum value of the transition amplitude is $\frac{1}{16} \approx 0.063$, 
while the 
minimum value is $\frac{1}{64}\approx 0.016$. As we increase $j$ (and so the number 
of spin-$\frac{1}{2}$ constituents), the transition amplitude decreases exponentially. Hence, 
we should pick $j$ to be as low as possible to have a feasible gate.

For most practical purposes, even the maximum transition amplitude is too low. To fix this, 
we can directly implement the repeat-until-success scheme proposed in Ref.~\cite{oi14}. 
In essence, by using two auxiliary states (such as $\ket{j, \pm (m \neq j)}$), one can define 
a binary tree of projections that guarantees that each of the four steps in our scheme 
happens with near certainty. 

Besides the transition amplitude, noise and decoherence are also present in our system. 
Although our reliance on holonomies offers some built-in protection \cite{fuentes05}, we can 
add an extra layer of security by realising that there is a natural link between discrete holonomies 
and quantum error correction. For clarity, we restrict our discussion to $j= \frac{3}{2}$. Observe 
that we can view our SCSs as logical qubits: $\ket{\frac{3}{2}} = \ket{\frac{1}{2}}^{\otimes 3} = 
\ket{0_{\rm L}}$ and $\ket{-\frac{3}{2}} = \ket{-\frac{1}{2}}^{\otimes 3} = \ket{1_{\rm L}}$. A 
bit flip code \cite{peres85} can be integrated directly into our scheme if we perform 
syndrome measurements and corrections after each projection onto a subspace. After 
completing the repeat-until-success procedure to get the qubit into the correct subspace, 
we perform error-correction measurements to correct any remaining single-qubit bit-flip errors. 
The syndrome measurements rotate with our states. This means a measurement of, say, 
$\sigma_z^i \otimes \sigma_z^j$ will need to be replaced by $\mathcal{D}_{\bf n}^i \sigma_z^i 
(\mathcal{D}_{\bf n}^i)^{\dagger} \otimes \mathcal{D}_{\bf n}^j \sigma_z^j 
(\mathcal{D}_{\bf n}^j)^{\dagger} =  \left( {\bf n}\cdot {\bm \sigma}^i \right) \otimes  
\left( {\bf n} \cdot {\bm \sigma}^j \right) $. This is advantageous because we only have 
to rotate before a gate measurement. Implementing the extra syndrome measurements 
and the repeat-until-success scheme does not introduce extra rotations. Furthermore,  
accurate rotations can be achieved experimentally because a rotated measurement of a 
state is equivalent to a collective rotation of the devices the read-out the spin-$\frac{1}{2}$
qubits.

It is also possible to implement the Shor code, which corrects arbitrary errors in single, 
physical qubits due to decoherence \cite{shor95}. So now, after the repeat-until-success 
procedure, we can correct arbitrary remaining single-qubit errors. We define the logical qubits 
as $\ket{0_{\rm L}} = \frac{1}{2\sqrt{2}} \left( \ket{\frac{3}{2}} + \ket{-\frac{3}{2}} \right)^{\otimes 3}$ 
and $\ket{1_{\rm L}} = \frac{1}{2\sqrt{2}} \left( \ket{\frac{3}{2}} - \ket{-\frac{3}{2}} \right)^{\otimes 3}$. 
Again, the syndrome measurements follow our states' rotation. Unlike the bit flip code, our 
logical qubit states are not the same as the SCSs. This means the exact rotation 
angles, given in Eq.~(\ref{eq:angles}), will be different. It is also nontrivial that our new 
overlap matrices still decompose into a scalar times a unitary matrix. To prove this, we 
abbreviate $e^{i\theta_a J_y} e^{-i\left( \phi_b - \phi_a \right) J_z} 
e^{-i \theta_b J_y} = U$. Sufficient conditions are that $\bra{1_{\rm L}} U \ket{0_{\rm L}} = 
- \bra{0_{\rm L}} U \ket{1_{\rm L}}^\ast$ and $\bra{1_{\rm L}} U \ket{1_{\rm L}} = 
\bra{0_{\rm L}} U \ket{0_{\rm L}}^\ast$. Since the three-groups of physical qubits are 
rotated by the same unitary $U$ we have that
    \begin{eqnarray}
         \bra{0_{\rm L}} U \ket{1_{\rm L}} & = & \left[ \left( \bra{\frac{3}{2}} + \bra{-\frac{3}{2}} \right) U 
         \left( \ket{\frac{3}{2}} - \ket{-\frac{3}{2}} \right) \right]^3 \nonumber \\
         & = & \left( R - R^\ast - S - S^\ast \right)^3, \nonumber\\
         \bra{1_{\rm L}} U \ket{0_{\rm L}} & = & \left( -R + R^\ast + S + S^\ast \right)^3 \nonumber \\
         & = & \left( -1 \right)^3 \left[ \left( R - R^\ast - S -S^\ast \right)^\ast \right]^3 \nonumber \\
         & = & -\bra{0_{\rm L}} U \ket{1_{\rm L}}^\ast.
    \end{eqnarray}
A similar calculation proves the diagonal elements. This shows that error-correcting 
codes and discrete holonomies can be merged into a single scheme.

We should mention that the repeat-until-success scheme and the syndrome measurements 
are two separate, distinct steps. It is not sufficient to combine the syndrome measurements 
and the projection measurements into a single step. This is because the different subspaces 
into which we project are not close together, so the probability of an error involving more than 
one qubit is high. These multi-qubit errors cannot be fixed with only the Shor code, which is 
why we also need the repeat-until-success scheme.

In line with this, it is the combination of the repeat-until-success scheme and the Shor code 
that makes our scheme fault-tolerant, up to errors in the syndrome measurements. Multi-qubit 
errors get fixed by the repeat-until-success scheme, and if any single-qubit errors remain, these 
get fixed by the Shor code. Only if the syndrome measurements of the Shor code go wrong 
do we get an error. In this sense, our scheme is as fault-tolerant as the Shor code.

\section{Zeno limit}
We conclude our discussion on discrete holonomic gates by verifying the continuous-path, 
or Zeno limit. It is a well-known result that taking the limit of infinitely many projective 
measurements `freezes' a quantum system, or drives it through different subspaces with 
effective certainty \cite{aharonov80}.

For our choice of SCSs, when we take the limit to dense 
measurements, we find that we can only implement a phase gate \cite{remark3}. To get 
a non-commuting holonomy with the SCSs, necessary for universality, we require a 
finite number of projections. In passing, we note that if $m = \pm \frac{1}{2}$ instead of 
$\pm j$ the holonomy would remain non-Abelian in the Zeno limit \cite{zee88}.

We take the Zeno limit for the $\mathcal{D}_z(\phi)$ gate. As there is some freedom in 
what measurement sequence to take, we pick $(0,0) \to \left( \frac{\pi}{2},0 \right) \to 
\left( \frac{\pi}{2},\varphi \right) \to (0,0)$ for simplicity. One can show that we expect a 
continuous-path holonomy given by $ U_{D} = \exp(ij \sigma_z \int_0^\varphi \mathrm{d}\phi)$.

\begin{figure}[htb]
    \subfloat[\label{fig:zenoA}]{\includegraphics[width= 4.1 cm]{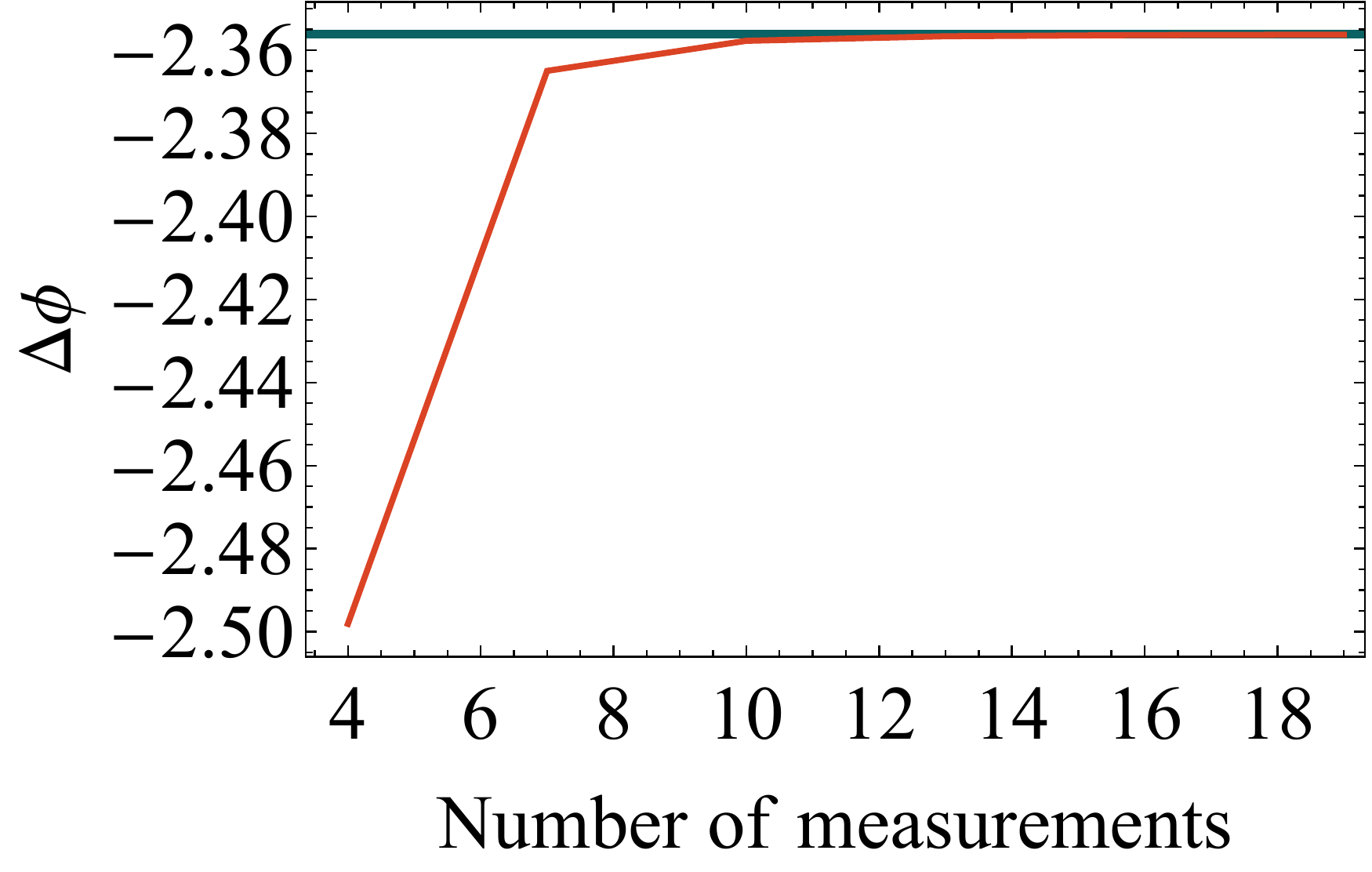}}
    \hfill
    \subfloat[\label{fig:zenoB}]{\includegraphics[width= 4.1 cm]{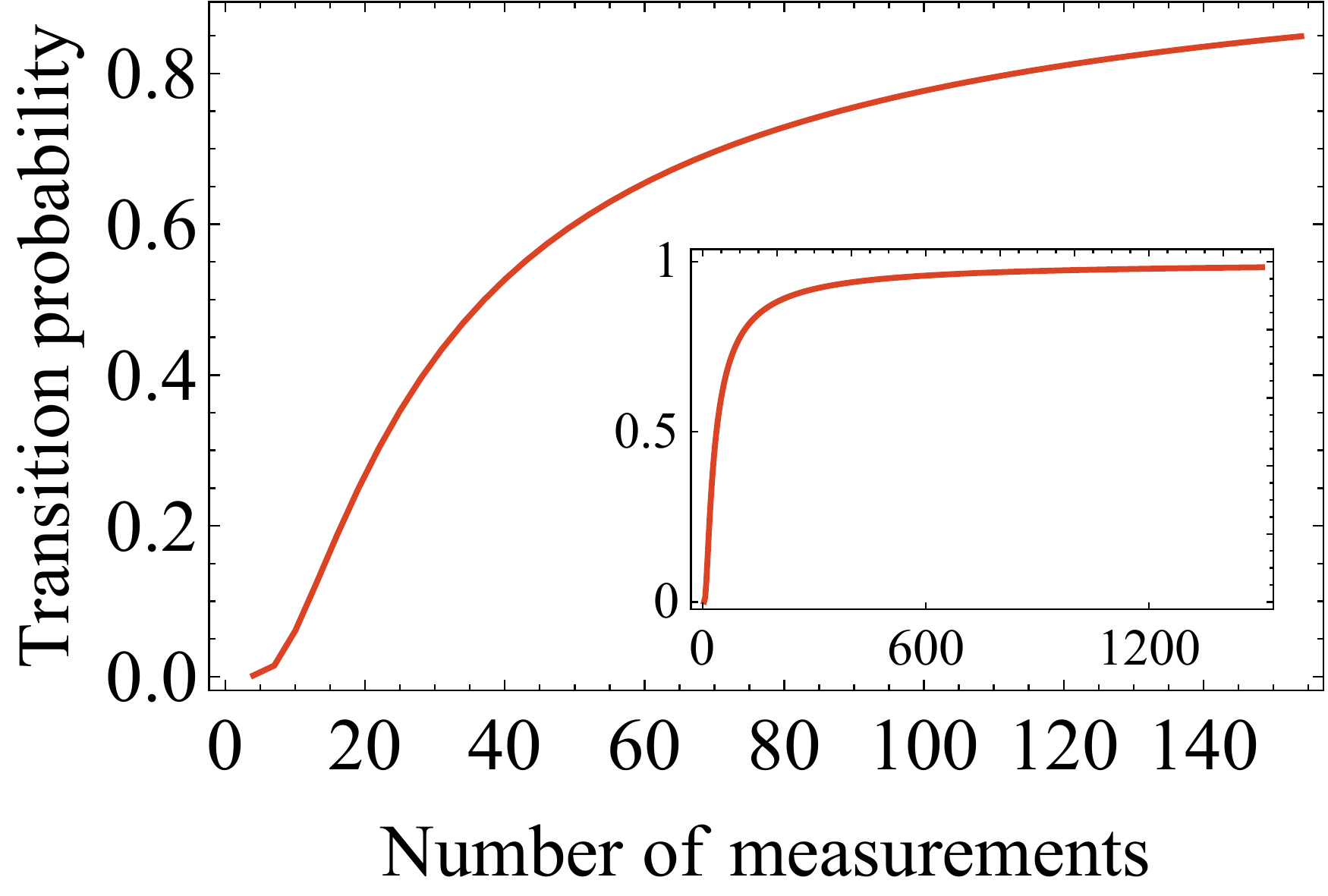}}
    \hfill
    \subfloat[\label{fig:zenoC}]{\includegraphics[width= 4.1 cm]{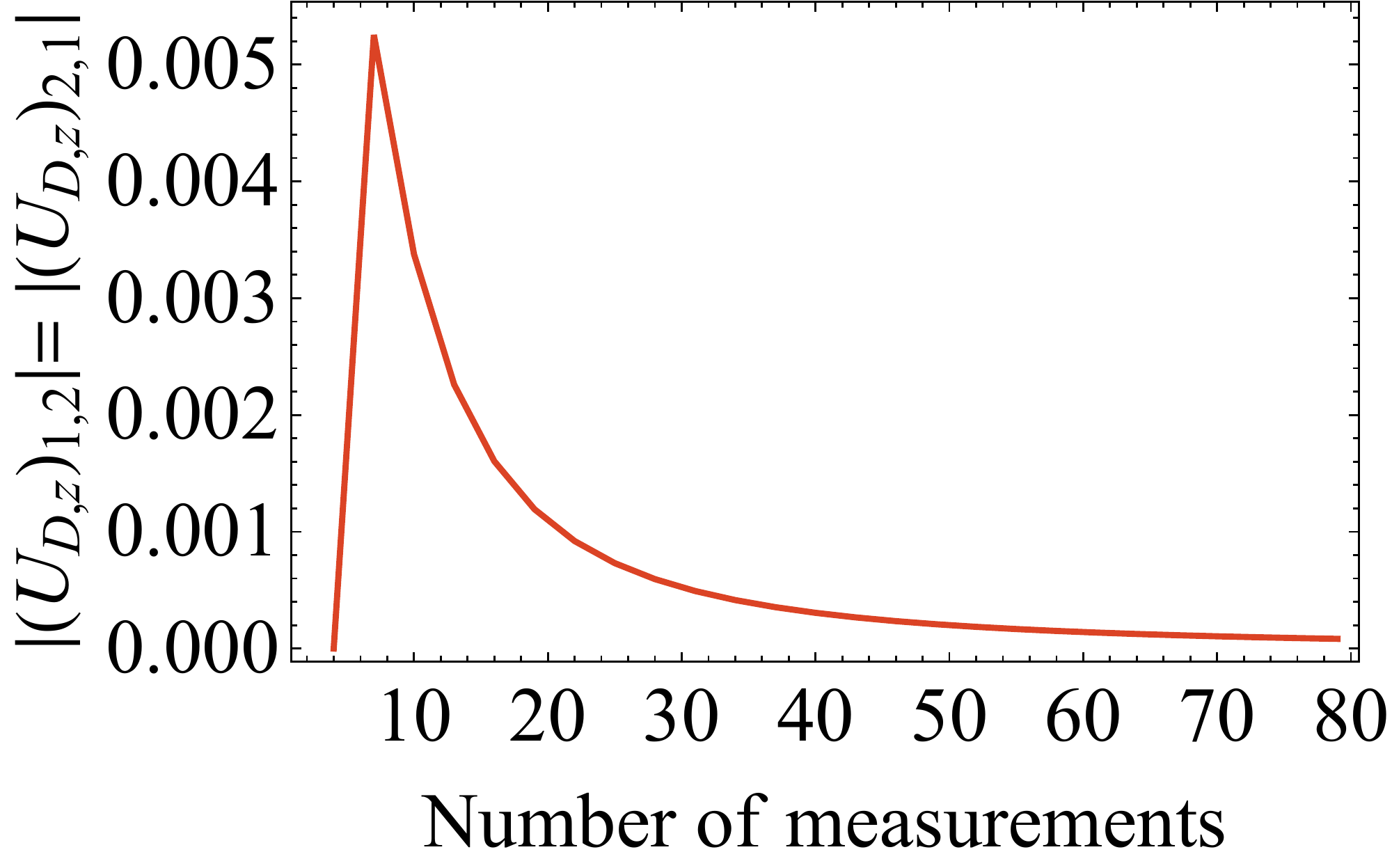}}
    \hfill
    \subfloat[\label{fig:zenoD}]{\includegraphics[width= 4.1 cm]{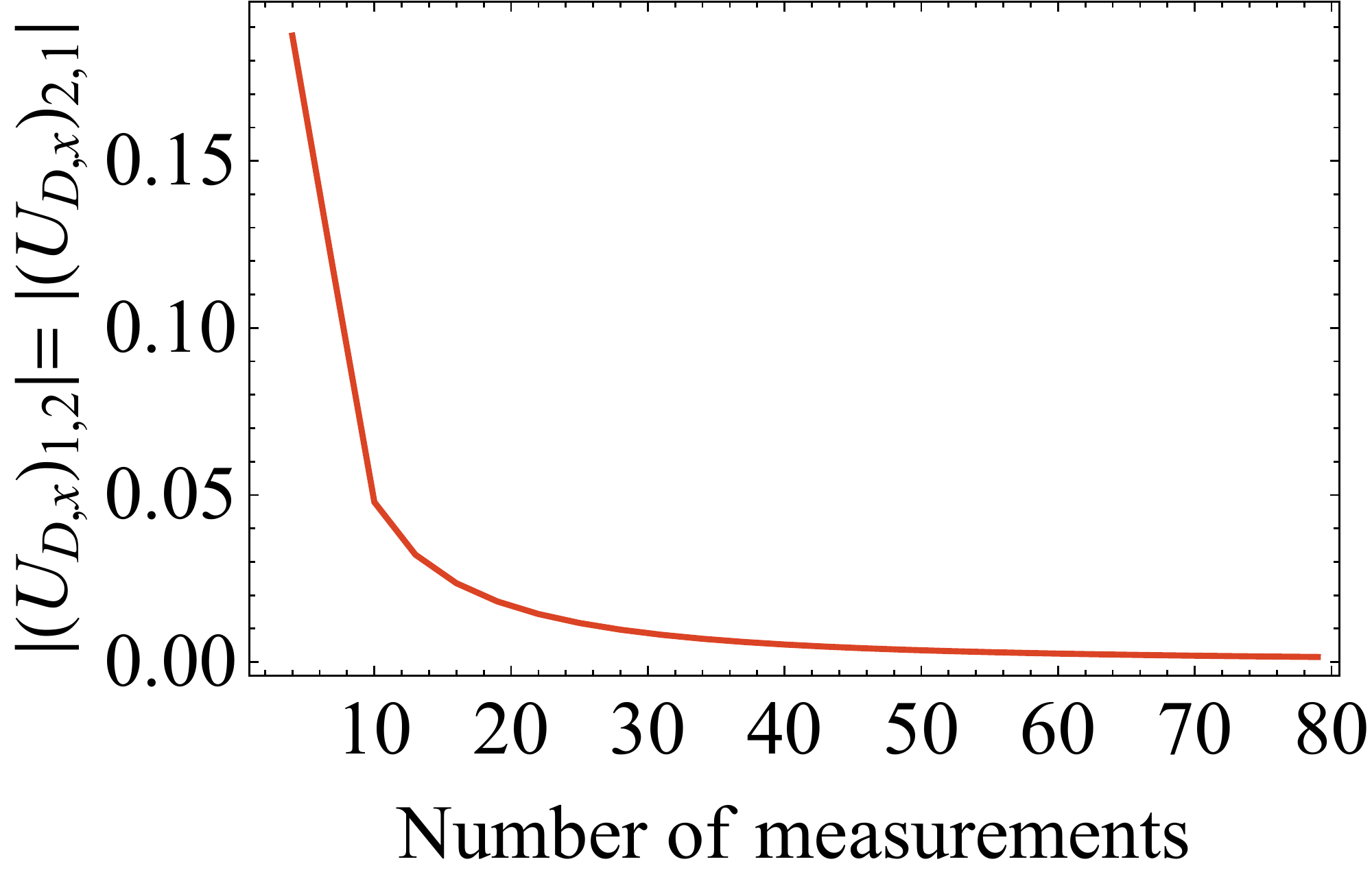}}
    \caption{The behaviour of the discrete holonomy $U_{D}$ for a subspace rotation of 
        $\varphi = \frac{\pi}{4}$ (using a Z-gate or an X-gate) when the number of measurements 
        is increased. (a) The argument of the relative phase of the diagonal elements of the 
        holonomy for the Z-gate. (b) The transition probability for an initial state $\frac{1}{\sqrt{2}} 
        \left( \ket{j} + \ket{-j} \right)$ for the Z-gate. The inset shows a zoomed-out version of 
        the larger plot. (c) The absolute value of the off-diagonal elements of the holonomy for 
        the Z-gate. (d) The absolute value of the off-diagonal elements of the holonomy for the X-gate.}
    \label{fig:zeno}
  \end{figure}

Suppose we have $j = \frac{3}{2}$ and rotate our subspace by $\varphi = \frac{\pi}{4}$. 
In the Zeno limit, our holonomy should be a phase shift with argument
$\Delta \phi = -\frac{3\pi}{4} \approx -2.35$ \cite{remark4}. If we choose the initial state 
$\ket{\psi} = \frac{1}{\sqrt{2}}\left( \ket{j} + \ket{-j} \right)$ 
(say), we can also calculate the transition probability. This probability should go to one 
as the number of measurements increases, because of the quantum Zeno effect. 
Furthermore, the off-diagonal elements of the holonomy $U_D$ should tend to zero. 
This means that increasing the number of measurements to increase the transition 
amplitude comes with the downside of changing the holonomy. That the off-diagonal 
matrix elements tend to zero also holds for holonomies that are initially non-Abelian, 
such as a rotation of $\frac{\pi}{4}$ around the $x$ axis. The argument of the relative 
phase shift of the diagonal elements should go to $-\frac{3\pi}{4}$ for $j=\frac{3}{2}$. 
Numerical calculations confirm this, as shown in Fig. \ref{fig:zeno}.

\section{Conclusions}  
In conclusion, we have demonstrated that discrete holonomic quantum computation can achieve 
universality. In particular, we have explicitly constructed quantum gates for spin coherent states, 
whereby rotation gates were achieved using a sequence of four projective measurements. We 
have further shown that it is possible to construct an entangling two-qubit gate. These results 
widen the scope of holonomic gates to include spin coherent states. Our scheme offers an extra 
layer of protection as we can readily integrate quantum error correction codes, such as Shor's 
code. The low transition rate for the projective measurements can be mitigated using a 
repeat-until-success scheme. This means that, at the cost of more measurements, the error 
correction code and repeat-until-success scheme can be combined, though this warrants a 
more detailed investigation. We also recover previously found results for continuous-path 
holonomic quantum computation by taking the limit to infinitely many measurements. Our 
study demonstrates a deep connection between quantum error correcting codes and discrete 
holonomies, and highlights the differences between the continuous and discrete realms. The 
proposed scheme provides a model for measurement-based quantum computation that 
combines the passive error resilience of holonomic quantum computation and active error 
correction techniques. Our work should be viewed as a proof-of-principle; our idea is not 
limited to the Shor code. It should be investigated to what extent other quantum codes can 
be merged with (discrete) holonomic quantum computation. For example, the $[[5,1,3]]$ code 
\cite{laflamme1996} could be used instead of the Shor code, or syndrome measurements could 
be supplemented with flag qubits \cite{chao2018} and gauge qubits \cite{campbell2017}. 

\vskip 0.3 cm 
E. S. acknowledges financial support from the Swedish Research Council (VR) through 
Grant No. 2017-03832.


\begin{thebibliography}{99}
\bibitem{remark0} Note that, besides mathematically, we should also physically interpret 
our state $\ket{j}$ to be a composite system of spin-$\frac{1}{2}$ particles. One should 
thus not think of $\ket{j}$ as (say) a spin-$j$ atom but as a composite of spin-$\frac{1}{2}$ 
qubits.
\bibitem{remark1} We stress that this requires $j$ to be a half-odd integer. 
If $j$ is an integer then $\kappa^{-1}$ ceases to be a number and becomes a positive semidefinite 
matrix instead, which is considerably more complicated when combined with the other overlap 
matrices in the sequence \cite{sjoqvist06}.
\bibitem{remark2} In the continuous-path case one can pick coordinates that remove one 
(but not both) of the singularities. The two patches that cover the system are related by a 
gauge transformation \cite{zee88}.
\bibitem{remark3} To see why, let us 
divide the total angles into $N$ steps, $\delta \theta = \theta/N$ and $\delta \phi = \phi/N$, 
and take the limit $N \to \infty$. Then, one of the overlap matrices $\left( \mathcal{F} 
(\theta_a+\delta\theta, \phi_a + \delta\phi) | \mathcal{F} (\theta_a,\phi_a) \right)$ contains 
(up to and including first order) $e^{i\left( \theta_a+\delta\theta \right) J_y} 
e^{i\left( \phi_a + \delta\phi \right) J_z}  e^{-i\phi_a J_z} e^{-i\theta_a J_y} = 
1 + i\delta\theta J_y + i\delta\phi \left( -\sin(\theta_a) J_x + \cos(\theta_a) J_z \right)$.
Since $j \geq 1$ the off-diagonal elements of the overlap matrix vanish and the diagonal 
elements become $1 \pm i j \cos \left( \theta_a \right) \delta\phi$. Therefore, in the limit 
$N \to \infty$ the total holonomy can only be a phase gate.
\bibitem{remark4} The appropriate way to take the Zeno limit is to increase 
the number of measurements along geodesics connecting the initial points. In taking the 
limit it does not matter from which $\phi$ coordinate we approach the $(0,0)$ subspace, 
as all subspaces at the north pole are connected by a $\mathrm{U}(1)$ gauge transformation 
and projectors are gauge invariant.
\end{thebibliography}
\end{document}